\documentclass[runningheads]{llncs}
\usepackage[T1]{fontenc}
\usepackage{graphicx}
\usepackage{amssymb}
\usepackage{tabularx}
\usepackage{bm} 
\usepackage[colorlinks, linkcolor=blue, anchorcolor=blue, citecolor=blue]{hyperref}
\usepackage{marvosym}
\begin{document}
\title{DSCENet: Dynamic Screening and Clinical-Enhanced Multimodal Fusion for MPNs Subtype Classification}
\titlerunning{DSCENet}
%

\author{Yuan Zhang\inst{1}\orcidID{0000-0002-0762-7514} \and Yaolei Qi\inst{1}\orcidID{0000-0002-8531-7386} \and Xiaoming Qi\inst{1}\orcidID{0000-0002-3238-2002} \and Yongyue Wei\inst{2}\orcidID{0000-0002-1132-1796} \and Guanyu Yang\inst{1,3}\orcidID{0000-0003-3704-1722} (\Letter)}

\authorrunning{Y. Zhang et al.}
%
\institute{Key Laboratory of New Generation Artificial Intelligence Technology and Its Interdisciplinary Applications (Southeast University), Ministry of Education \\ \email{yang.list@seu.edu.cn} \and
Center for Public Health and Epidemic Preparedness Response, Peking University \and
Jiangsu Provincial Joint International Research Laboratory of Medical Information Processing, Southeast University}

\maketitle             
\begin{abstract}
The precise subtype classification of myeloproliferative neoplasms (MPNs) based on multimodal information, which assists clinicians in diagnosis and long-term treatment plans, is of great clinical significance. However, it remains a great challenging task due to the lack of diagnostic representativeness for local patches and the absence of diagnostic-relevant features from a single modality. In this paper, we propose a Dynamic Screening and Clinical-Enhanced Network (DSCENet) for the subtype classification of MPNs on the multimodal fusion of whole slide images (WSIs) and clinical information. (1) A dynamic screening module is proposed to flexibly adapt the feature learning of local patches, reducing the interference of irrelevant features and enhancing their diagnostic representativeness. (2) A clinical-enhanced fusion module is proposed to integrate clinical indicators to explore complementary features across modalities, providing comprehensive diagnostic information. Our approach has been validated on the real clinical data, achieving an increase of 7.91\%  AUC and 16.89\% accuracy compared with the previous state-of-the-art (SOTA) methods. The code is available at \href{https://github.com/yuanzhang7/DSCENet}{https://github.com/yuanzhang7/DSCENet}.
\end{abstract}

\section{Introduction}
The precise subtype classification of MPNs based on multimodal information is crucial \cite{arber2022internationalMPN}. MPNs are a heterogeneous group of hematologic malignancies with marked disparities in the progression of distinct subtypes, including polycythemia vera (PV), essential thrombocythemia (ET), prefibrotic/early primary myelofibrosis (PrePMF), and primary myelofibrosis (PMF) \cite{arber20162016MPN,baumeister2021progressionMPN}. For example, PMF has an increased risk for acute myelogenous leukemia transformation \cite{yogarajah2017leukemic}, while ET and PV give rise to thrombotic and bleeding complications \cite{tefferi2017polycythemia}. However, relying solely on one modality, such as pathological images, yields limited insights \cite{barbui2016myeloproliferative}. The misclassification of subtypes may lead to inappropriate therapy, affecting treatment efficacy and patient survival \cite{arber20162016MPN}. Hence, accurate multimodal MPNs classification holds urgent clinical demands.

\begin{figure}[tp]
\includegraphics[width=\textwidth]{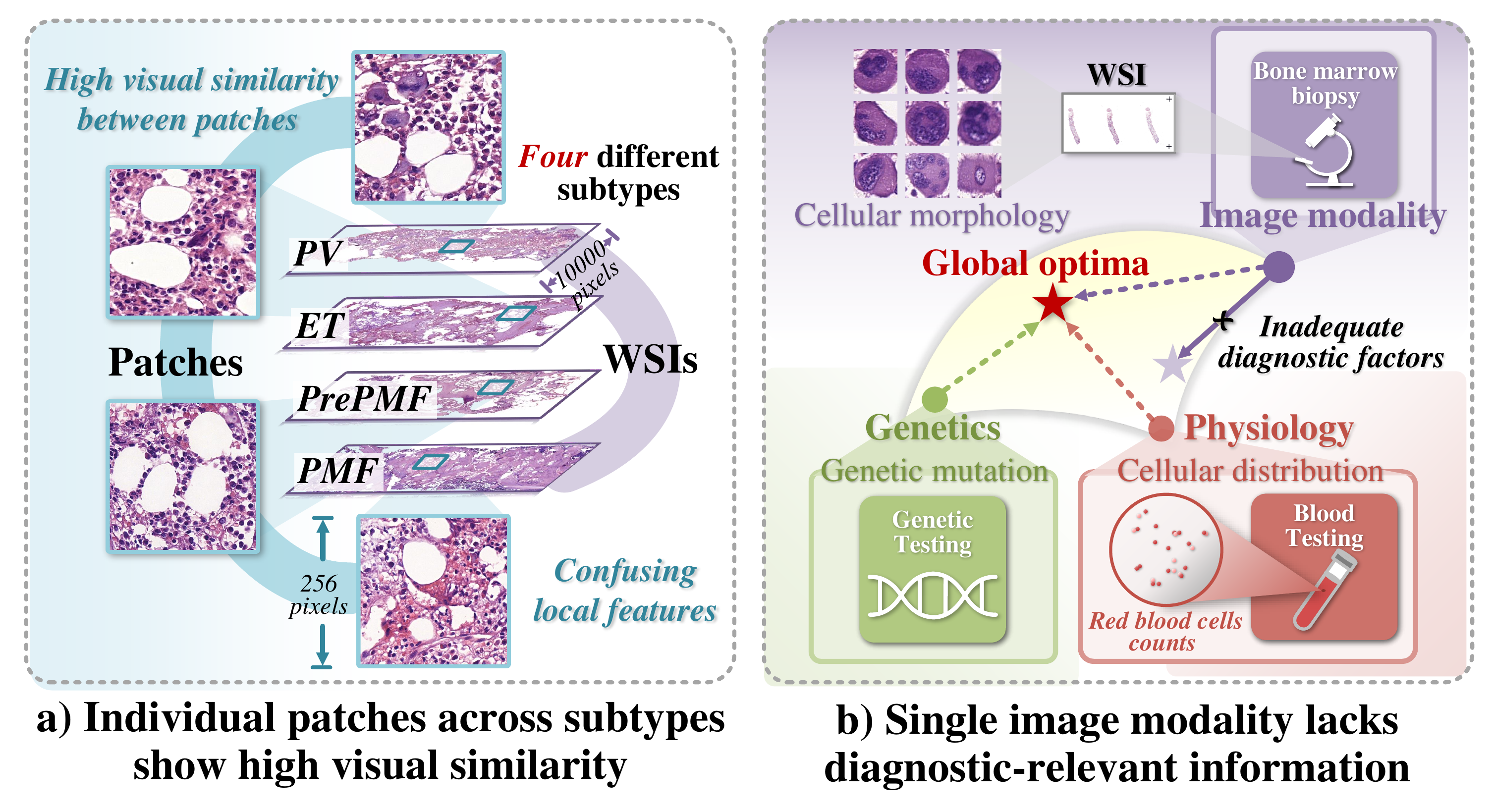}
\caption{ a) Challenge 1: Individual patches show high visual similarity across four subtypes, leading to confusing local features. b) Challenge 2: Single image modality only provides morphological information, lacking other diagnostic-related features.} 
\label{fig1}
\end{figure}

Pathology analysis on bone marrow biopsy whole slide images (WSIs) serves as the clinical gold standard for MPNs diagnosis \cite{ryou2023quantitativeMPN,arber2022internationalMPN}, posing distinct challenges due to a unique paradigm of proliferative disorders. \textbf{Challenge 1: Individual patches lack diagnostic representativeness.} Due to the dispersed distribution and dense proliferation of blood cells, MPNs lack fixed lesion areas, causing individual patches to exhibit high visual similarity across four subtypes as Fig.~\ref{fig1}. a) shown. Constraining tens of thousands of pixels of the WSI with a single category label is inadequate for capturing the intricate features and diversity inherent. The difficulty arises when local patches with similar textures struggle to represent different categories in the feature space, potentially leading to increased ambiguity and failure to encapsulate representative features. Hence, the effective selection of local patches from the perspective of holistic semantics is crucial for enhancing the model's ability to identify diagnostically relevant features. \textbf{Challenge 2: Single image modality lacks diagnostic-relevant features.} The diagnosis of MPNs relies on the comprehensive consideration of multi-modal diagnostic information such as morphology, physiology, and genetics \cite{arber2022internationalMPN}. However, the single modality of images only focuses on morphological features and lacks other diagnostic-relevant features, such as the driver mutations from genetics and cell distributions from physiology, leading to the model being trapped in local optima as Fig.~\ref{fig1}. b) shown. Therefore, incorporating clinical information guides the model in better learning the correlations and complementarities among different modalities, improving classification performance. Overall, the aforementioned MPN-related studies have feature extraction in isolation and have not facilitated multimodal fusion from the feature learning process.

Recently, while deep learning algorithms have demonstrated outstanding capabilities in pathological image diagnosis \cite{he2024foundation,FedSODA}, the classification of hematological disorders like MPNs is still in its early stages \cite{elsayed2023applications}. Existing methods can be broadly classified into three types: 1) Pathology image-based approach, which often necessitates experts to perform cell-level annotations for the morphological analysis\cite{krichevsky2023unetmpncell,kimura2021automated}. \cite{bruck2021machine,sirinukunwattana2020artificial} extract features from WSIs based on the fingerprinting of megakaryocytes, requiring extensive cell annotation. \cite{yusof2022hyperparameter} employs CNN models to classify patches directly but overlooks global contextual information. 2) Clinical information-based approach, which relies on indicators such as cell counts for diagnosis. \cite{meggendorfer2017deepclinic} focuses on clinical indicators but faces limitations from the quality and completeness of available data. 3) Multimodal integration approach, which combines clinical data with pathology images. \cite{wang2023development} integrates predictive probabilities from images and clinical data using statistical regression, but lacks interactive fusion at the feature learning level.

Multimodal fusion enhances comprehensive diagnostic information for MPNs, guiding the model to simultaneously learn features from different modalities, thereby exploiting cross-modal complementarity and improving classification accuracy \cite{song2023artificial,cui2023mulmodalreview}. To address challenge 1, we propose a dynamic screening (DS) module to filter the feature relevance of local patches, constraining the potential distribution differences to calibrate features, thereby boosting the model's representation. To address challenge 2, we propose a clinical-enhanced fusion (CF) module which introduces clinical information as additional guidance for image features learning. Given the dimensional disparity between modalities, our CF explores complementary features relevant to diagnosis in cross-modal learning, facilitating comprehensive diagnosis.

In this paper, we propose the multimodal fusion framework, dynamic screening and clinical-enhanced network (DSCENet), aiming at MPNs subtype classification. DSCENet capably explores complementary information across modalities, heightening the model's representative capacity and diagnostic accuracy. Our contributions are as follows:
\begin{itemize}
    \item 
    To the best of our knowledge, DSCENet achieves MPNs subtype classification based on multimodal fusion for the first time, obtaining significant improvements and contributing to reliable treatment plans.
    \item 
    We propose a dynamic screening module to flexibly select the local patches, aiding the model in capturing pivotal features and improving representation capacity and generalization performance.
    \item 
    We propose a novel clinical-enhanced fusion module to mine the diagnostic-relevant complementary features across modalities by using clinical features as additional guidance, facilitating comprehensive diagnostic accuracy.
     \item 
     Our DSCENet outperforms the SOTA methods on the real clinical dataset consisting of 383 cases with an accuracy of 83.12\% and an AUC of 96.43\%.
\end{itemize}

\section{Methodology}

\begin{figure}[tp]
\includegraphics[width=\textwidth]{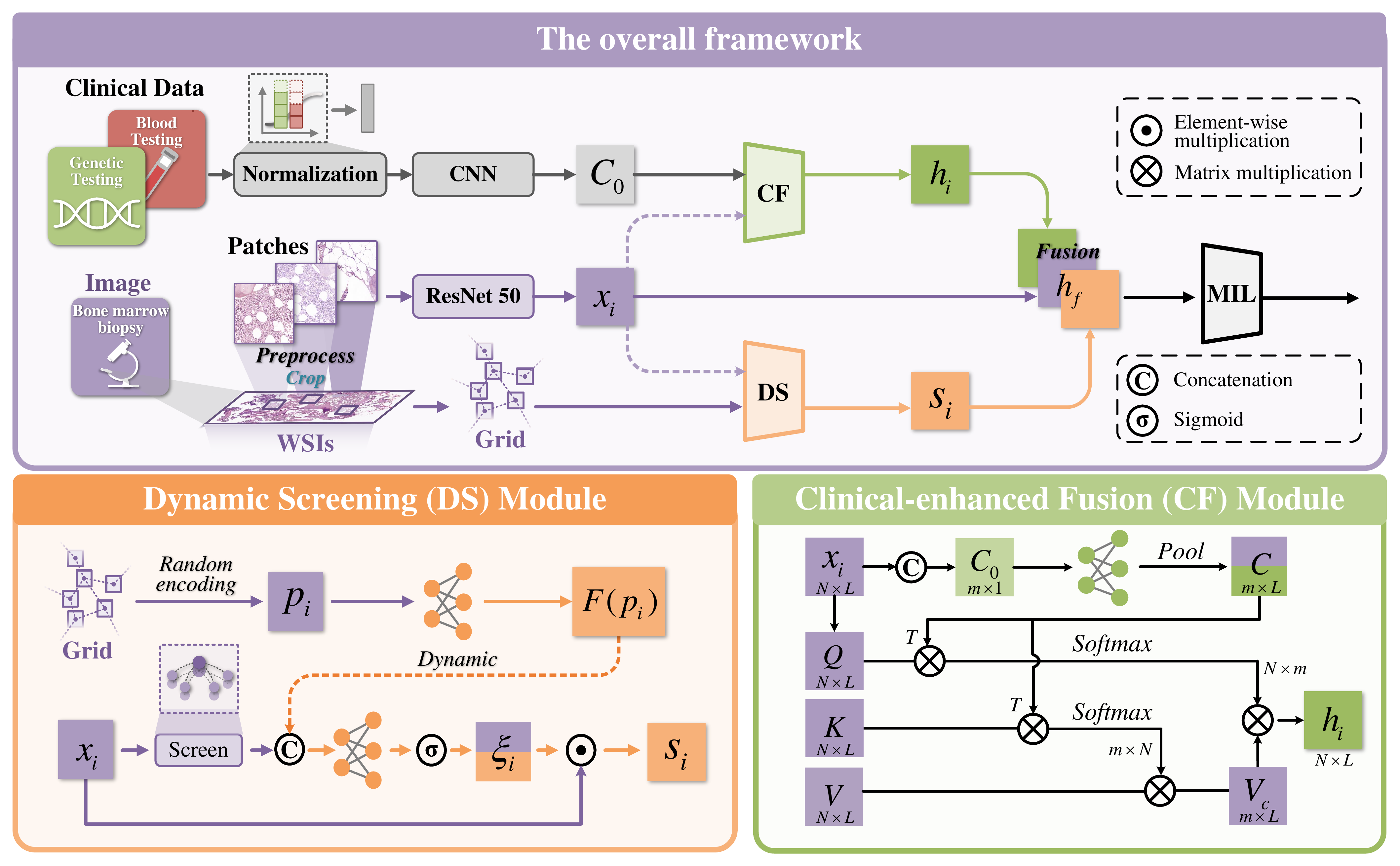}
\caption{ The Framework of our DSCENet. The DS module aims to dynamically screen the local patches for better feature representation and clinical-enhanced fusion module explores the complementary diagnostic information. } 
\label{fig2}
\end{figure}

Fig.~\ref{fig2} illustrates the overall framework of our method to integrate WSIs and clinical data, which consists of two main modules. The dynamic screening module aims to flexibly screen the feature learning of local patches, improving the representative capacity of WSIs. The clinical-enhanced fusion module intends to integrate clinical indicators to guide the exploration of complementary features across modalities, offering comprehensive diagnostic information.

\subsection{Dynamic Screening Module}
\textbf{Dynamic random encoding of the grid.} 
Our approach is dedicated to enhancing the representation of local patches in the holistic context of WSIs, i.e., $B = \{x_{i}\}$ denotes the bag of features per WSI, where $i$ is the index of the cropped patches. The feature embedding of patch $x_{i} \in\ \mathbb{R}^{N \times L}$ is extracted by the pre-trained ResNet 50 \cite{lu2021clam}, where $N$ is the number of patches and $L$ is the feature dimension. Instead of the absolute position encoding, we dynamically random encode the sequences of patches as the random grid $p_{i} \in\ \mathbb{R}^{N}$. 
Then, two fully-connected layers with ReLU activation $\mathcal{F}(\cdot)$ are used to realize dynamic enhancement encoding. The dynamic random encoding of the grid $\mathcal{F}(p_i)$ efficiently adapts to the varying number of patches within each WSI, thereby aiding the model in capturing the overall structure and patterns in the sequence.

\textbf{Holistic screening of patch feature.} Unlike the top-ranking tiles \cite{campanella2019clinical} that discards subsequent patches, the dispersed distribution of MPNs requires retraining the holistic feature information through dynamic screening. To capture the holistic semantics, we also perform global averaging $\mathcal{G}(\cdot)$ on ${x_{i,j}}$ to save the computational cost. Following by a full-connection layer and the gating sigmoid activation for augmenting embedding by the textural knowledge, the selection descriptor of patch feature $\xi_i$ is then given by: 
\begin{equation}
\xi_i = sigmoid ( \mathcal{G}(x_{i})  \oplus  \mathcal{F}(p_{i}) ),
\end{equation}
where $\oplus$ denotes concatenation. 

Finally, we also implement a residual design to alleviate the adverse impacts of inaccurate selections. The dynamic selected feature $s_{i}$ is then given by:
\begin{equation}
s_{i} = x_{i} \otimes \xi_{i} + x_{i},
\end{equation}
where $\otimes$ denotes pixel-wise multiplication. As the selection descriptor $s_{i}$ are mutually independent and differ among local patches, they generate different augmented versions to calibrate the feature representation learning and eventually adapt to the holistic semantics.

\subsection{Clinical-enhanced Fusion Module} 
Given the dimensional disparity between WSIs and clinical data in MPNs diagnosis, multimodal fusion methods like alignment are suboptimal \cite{cui2023mulmodalreview}. Inspired by \cite{han2023agent}, which applies agent tokens to enhance efficiency, we design the clinical-enhanced fusion module to mine the diagnostic-relevant complementary features by introducing clinical features as additional guidance between clinical-enhanced query and clinical-enhanced value block. 

\textbf{Clinical query block.}
Clinical indicators $C_o \in \mathbb{R}^{M}$ includes genetics, reflecting the correlation between gene mutations and specific subtypes \cite{barbui2018MPNneedgene}, and physiology, describing dynamic physiological metabolic distributions \cite{arber2022internationalMPN}, where $m$ is the number of clinical indicators. Clinical-enhanced query $ C \in \mathbb{R}^{M \times L} $ is derived from the concatenation of clinical feature and image feature followed by fully connected layers and ReLU activation and scale pooling. The image feature $Q = x_iW_Q, K = x_iW_K, V = x_iW_V$, where $W_{Q/K/V}$ denote projection metrices and $d_k$ is the dimension. Clinical query block performs attention calculations between $C, K, V$ to query the focus of clinical attention from images (Equ. \ref{eq:attention1}).

\begin{equation}
V_c=Attention(C,K,V) = softmax(\frac{CK^T}{\sqrt{d_k}})V,
\label{eq:attention1}
\end{equation}

\begin{equation}
h_i=Attention(Q, C, V_c) = softmax(\frac{QC^T}{\sqrt{d_k}})V_c,
\label{eq:attention2}
\end{equation}

\textbf{Clinical-enhanced shared block.} To obtain complementary features $h_i$ cross modalities, it is essential to perform the second attention with image feature as Equ. \ref{eq:attention2}. Clinical-enhanced shared block achieves bidirectional attention in the shared feature space of both image and clinical modalities, offering comprehensive diagnostic insights and ultimately enhancing diagnostic precision.

Finally, the comprehensive feature $h_f$ is obtained after combining dynamical screening and clinical-enhanced fused feature: $h_f= s_i\oplus h_i$. Then we adopt the widely-used multiple instance classifier \cite{maron1997MIL} and Softmax for classification.

\section{Experiments}
\subsection{Experimental Settings} 
\subsubsection{Dataset.} 
To validate the effectiveness of our method, we conducted a multicenter study on MPNs data. 383 cases of MPNs are collected from 3 independent medical centers, including 81 PV, 126 ET, 88 PrePMF, and 88 PMF cases. Each patient tasks one WSI paired with its corresponding clinical information with no missing data, including gender, age, and blood test report (hemoglobin, white blood cell, red cell mass, hematocrit, platelet count) and genetic mutations (JAK2, MPL, CALR). The WSIs are stained by Hematoxylin and Eosin and scanned at 0.2 m/pixel using PANNORAMIC SCAN 150.

\subsubsection{Implementation Details.} 
We preprocess WSIs into non-overlapping patches measuring $256 \times 256$ pixels at 20 $\times$ magnification and use a pre-trained ResNet-50 \cite{lu2021clam} to extract the 1024-dimensional morphological features. All experiments are conducted on the two NVIDIA GeForce RTX 4090 (24 GB). The datasets are randomly split into training, validation, and testing sets with a ratio of 6: 2: 2. A total of 200 training epochs are conducted with a learning rate of $1 \times 10^{-4}$ using the Adam optimizer and cross-entropy loss.

\subsubsection{Evaluation Metric.} 
We perform comparative experiments and ablation studies to demonstrate the advantages of our methods. The area under the curve (AUC), Receiver Operating Characteristic curves (ROC), accuracy(ACC), precision, recall, and F1 score (F1) are employed to evaluate the performance. 

\subsection{Result and Analysis}
We performed comparative experiments and ablation studies to demonstrate the advantages of our proposed framework. The classical classification network CLAM \cite{lu2021clam}, Transmil \cite{shao2021transmil}, and DSMIL \cite{li2021DSMIL} are compared to validate the accuracy for the single image modality. We also compare with SNN \cite{klambauer2017SNN} for the single modality of clinical data. To validate the performance of multimodal fusion, we compared the AMIL \cite{ilse2018attentionmil}, MCAT \cite{chen2021MCAT} and CMTA \cite{zhou2023CMTA}. These models are trained on the same dataset and evaluated by the above metrics. For a fair comparison, the same CNN extractor and cross-entropy loss are adopted for all methods.  

\begin{table}[h]
\centering
\caption{Comparisons between our proposed method and other SOTA approaches (\%). $concat$ is concatenation fusion and $bilinear$ is bilinear pooling fusion.}
\begin{tabular}{ccc|ccccc}
\hline
                        & \textbf{WSIs}          & \textbf{Clinical data} & \textbf{ACC} & \textbf{AUC}     & \textbf{Precision} & \textbf{Recall}  & \textbf{F1} \\ \hline
CLAM\_SB \cite{lu2021clam}       & $\checkmark$ &               & 32.47  & 51.30 & 8.12   & 25.00 & 12.25  \\ 
CLAM\_MB \cite{lu2021clam}       & $\checkmark$ &               & 64.94  & 83.36 & 64.13   & 63.10 & 63.23  \\ 
DSMIL \cite{li2021DSMIL}       & $\checkmark$ &               & 53.25  & 78.86 & 72.66   & 60.84 & 54.95  \\ 
Transmil \cite{shao2021transmil}       & $\checkmark$ &               & 64.94  & 85.52 & 62.73   & 62.36 & 62.00  \\ 
SNN \cite{klambauer2017SNN}            &              & $\checkmark$  & 63.64  & 82.61 & 48.34   & 63.44 & 54.80  \\ \hline
AMIL\_bilinear \cite{ilse2018attentionmil} & $\checkmark$ & $\checkmark$  & 48.05  & 76.03 & 50.79   & 46.99 & 45.40  \\ 
AMIL\_concat \cite{ilse2018attentionmil}   & $\checkmark$ & $\checkmark$  & 58.44  & 88.52 & 53.02   & 64.93 & 56.21  \\ 
CMTA \cite{zhou2023CMTA}           & $\checkmark$ & $\checkmark$  & 58.44  & 87.19 & 62.96   & 62.38 & 57.13  \\ 
MCAT \cite{chen2021MCAT}           & $\checkmark$ & $\checkmark$  & 66.23  & 86.28 & 63.80   & 66.83 & 64.58  \\ 
\textbf{Ours} & $\checkmark$ & $\checkmark$ & \textbf{83.12} & \textbf{96.43} & \textbf{83.32} & \textbf{83.15} & \textbf{83.02} \\ \hline
\end{tabular}
\label{tab:table_result}
\end{table}

\subsubsection{Quantitative Evaluation.}
In the quantitative evaluation, our method achieve the best classification performance with 83.12\% accuracy, 96.73\% AUC, 83.32\% precision, 83.15\% recall, and 83.02\% F1 score in the Table \ref{tab:table_result}. Our method outperforms the second-best fusion model (MCAT) by 16.89\% in AUC and 7.91\% in accuracy and achieves an 18.18\% higher accuracy and 10.91\% higher AUC compared to the single-image modality model (Transmil). On the one hand, single-modality approaches utilizing either clinical information or WSIs have demonstrated moderate classification performance, indicating that individual modalities indeed contain specific diagnostic-relevant features. On the other hand, other modality fusion methods ignore the huge dimensional difference or design for prognostic tasks, which may not be applicable in the unique diagnostic scenarios of MPNs, thus failing to achieve the ideal alignment between two image and non-image feature spaces. However, our method employs a monitoring process augmented with additional clinical information, aiding the network in learning better features and improving classification performance.

\begin{figure}[h]
\includegraphics[width=\textwidth]{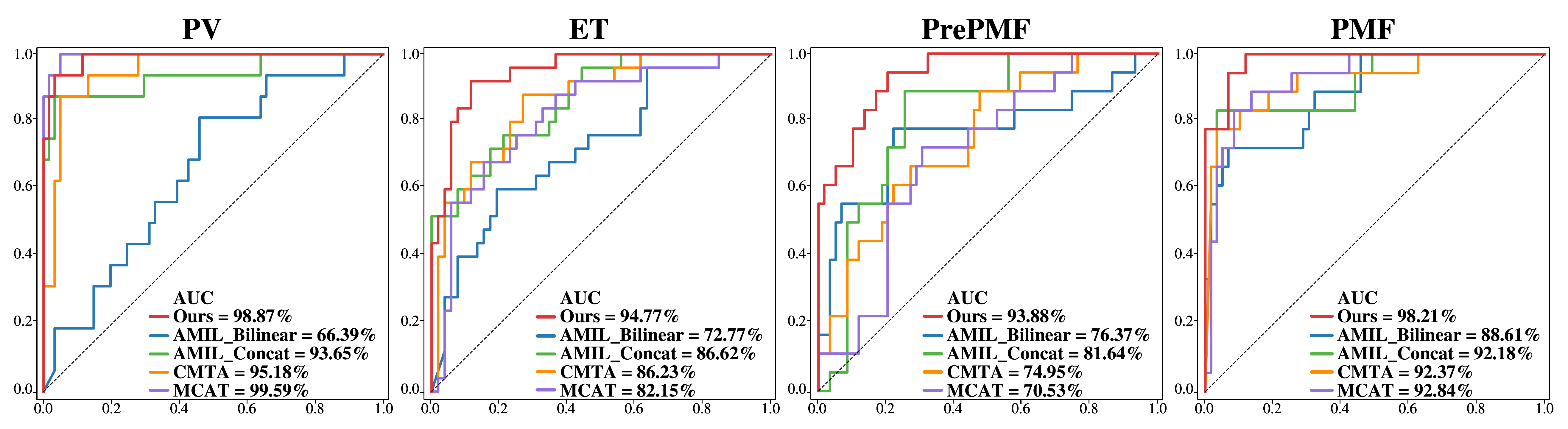}
\caption{ROC curves for different methods on PV, ET, PrePMF, and PMF.} 
\label{fig3}
\end{figure}

\subsubsection{Qualitative Evaluation.}
In the subtype classification task of MPNs, our model performs significantly superior for four subtypes on the ROC curves. The AUC values of our model are 98.87\%, 94.77\%, 93.88\%, and 98.21\% for PV, ET, PrePMF, and PMF, respectively, as depicted in Fig.~\ref{fig3}. Our model also surpasses the second-highest model by 8.54\%, 18.93\%, and 5.84\% in terms of AUC for ET, PrePMF, and PMF, respectively. Our model demonstrates well-balanced classification performance and achieves the top AUC value of 96.43\%, indicating effective fusion of clinical data with images at the cross-modal feature space, contributing to complementary diagnostic information.

\begin{table}[t]
\centering
\caption{Ablation results of our method (\%). DS is the dynamic screening module, and CF is the clinical-enhanced fusion module.}
\begin{tabular}{ccc|ccccc}
\hline
                                  &  &  & \textbf{ACC} & \textbf{AUC}     & \textbf{Precision} & \textbf{Recall}  & \textbf{F1} \\ \hline
w/o DS, w/o CF                       &                   &                          & 63.64  & 81.56 & 72.66   & 60.84 & 54.95  \\ 
w/o CF       &      &                          & 76.62  & 92.51 & 77.28   & 77.16 & 74.95  \\ 
w/o DS &                   &              & 76.62  & 93.90 & 75.35   & 76.38 & 75.09  \\ 
\textbf{Ours}                     &      &             & \textbf{83.12}  & \textbf{96.43} & \textbf{83.32}   & \textbf{83.15} & \textbf{83.02}  \\ \hline
\end{tabular}
\label{tab:table_ablation}
\end{table}

\begin{figure}[b]
\includegraphics[width=\textwidth]{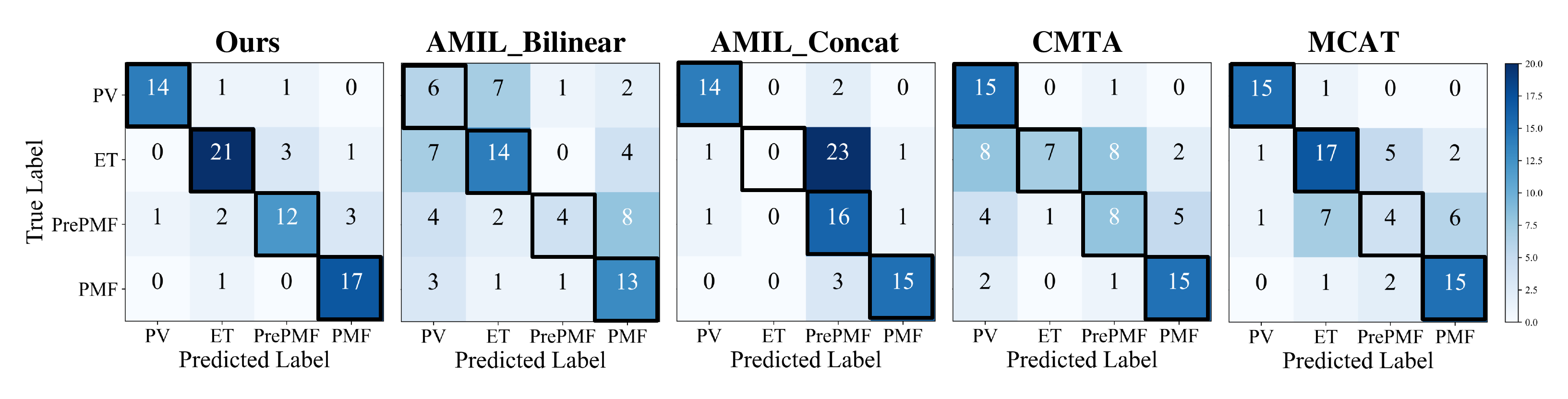}
\caption{Confusion matrices on the testing set, with true labels on the vertical axis and predictions on the horizontal axis. The deeper colors indicate higher accuracy.} 
\label{fig4}
\end{figure}

\subsubsection{Ablation Study.}
We evaluated the importance of monitoring and fusion modules by ablation experiments (Table \ref{tab:table_ablation}). The DS module brings significant improvement for classification, suggesting effective filtering of redundant features from patches. The CF module also improved the AUC by 12.34\%, indicating that clinical data with images enhances the representation of diagnostic relevance.

\subsubsection{Model analysis of confusion matrices.}
The confusion matrices of our model on the testing dataset of 77 samples demonstrate the distinguished classification performance of our model, achieving an accuracy of 83.11\% in Fig.~\ref{fig4}. It is evident that other methods are prone to misclassify ET as PrePMF or PV, whereas our method improves the accuracy of ET while maintaining the correctness of others. ET has dense cellular features, suggesting that our method can effectively enhance the local features with stronger diagnostic discrimination.

\section{Conclusion}
We propose a DSCEnet for the multimodal subtype classification of MPNs for the first time. This method effectively integrates pathological images and clinical data, achieving the optimal classification performance. Furthermore, we utilize a dynamic screening module to select patch features for better representation and a clinical-enhanced fusion module to explore diagnostic-relevant features for comprehensive diagnoses. We have pioneered a promising future for multimodal diagnosis of MPNs, offering vast prospects in diagnosis and treatment.

\begin{credits}
\subsubsection{\ackname} We thank the Big Data Computing Center of Southeast University for providing the facility support. We also thank Jiangsu Provincial People's Hospital, Peking Union Medical College Hospital, and the First Affiliated Hospital of Soochow University for providing the data.

\subsubsection{\discintname}
The authors have no competing interests to declare that are relevant to the content of this article.
\end{credits}

%
\bibliographystyle{splncs04}
\bibliography{selfbib}

\end{document}


\section{Experimental Settings}

\begin{table}[]
\centering
\caption{The distribution of Dataset Split}
\begin{tabular}{|c|c|c|c|c|c|}
\hline
                            & PV & ET & PrePMF & PMF & sum \\ \hline
\textbf{Train dataset}      & 49 & 76 & 52     & 52  & 229 \\ \hline
\textbf{Validation dataset} & 16 & 25 & 18     & 18  & 77  \\ \hline
\textbf{Test dataset}       & 16 & 25 & 18     & 18  & 77  \\ \hline
\end{tabular}
\end{table}

\section{Experimental details}
The clinical data in this paper utilize Min-Max Normalization, including m items of information,
For genetic mutation, ‘positive’ is encoded as 1 and ‘negative’ is encoded as 0.

\section{Details of the Evaluation Metrics}
More detailed information of the evaluation metrics is provided here, where the ground truth label is denoted as $G$, and the predicted label is denoted as $P$.

\textbf{Accuracy:}  Accuracy is calculated as the number of correct predictions divided by the total number of predictions made, expressed as a percentage. Accuracy essentially tells the proportion of correctly classified instances out of all instances evaluated. It formulates as Equ.~\ref{equ:accuracy}.

\begin{equation}
	\label{equ:accuracy}
	Accuracy = \frac{TP+TN}{TP+TN+FP+FN}
\end{equation}

\textbf{Precision:}  Precision is calculated as the number of true positive predictions divided by the sum of true positive and false positive predictions. Precision is particularly useful when the cost of false positives is high. It helps to understand how precise the model is in correctly identifying positive instances. It formulates as Equ.~\ref{equ:Precision}.

\begin{equation}
	\label{equ:Precision}
	Precision = \frac{TP}{TP+FP}
\end{equation}

\textbf{Recall:}  Recall is a measure of the ability of a model to correctly identify all relevant instances, or the true positives, within a dataset. It's calculated as the number of true positive predictions divided by the sum of true positive predictions and false negative predictions. Recall is important when the cost of missing relevant instances, or false negatives, is high. It helps to evaluate how well a model can capture all the relevant information within a dataset. It formulates as Equ.~\ref{equ:Recall}.

\begin{equation}
	\label{equ:Recall}
	Recall = \frac{TP}{TP+FN}
\end{equation}

\textbf{F1:}  F1 is is a metric that combines both precision and recall to provide a single measure of a model's performance. The F1 score balances between precision and recall, making it a useful metric for evaluating models when there's an uneven class distribution or when false positives and false negatives have different costs. A higher F1 score indicates better overall performance of the model. It formulates as Equ.~\ref{equ:F1}.

\begin{equation}
	\label{equ:F1}
	F1 = \frac{2*Precision*Recall}{Precision+Recall} = \frac{2*TP}{2*TP+FP+FN}
\end{equation}

\textbf{Receiver Operating Characteristic (ROC):}  ROC curve is a graphical representation of the trade-off between the true positive rate (sensitivity) and the false positive rate (1 - specificity) for a binary classification model, where the sensitivity and specificity values are formulates as Equ.~\ref{equ:sensitivity} and Equ.~\ref{equ:specificity}. It plots the true positive rate (TPR) against the false positive rate (FPR) at various threshold settings. It formulates as Equ.~\ref{equ:roc}. 

\begin{equation}
\label{equ:sensitivity}
	Sensitivity = Recall = \frac{TP}{TP+FN}
\end{equation}

\begin{equation}
	\label{equ:specificity}
	Specificity = \frac{TN}{FP+TN}	
\end{equation}

\begin{equation}
	\label{equ:roc}
	ROC = ???
\end{equation}

\textbf{Area Under the Curve (AUC):} AUC refers to the area under the ROC curve. It formulates as Equ.~\ref{equ:auc}. 

\begin{equation}
	\label{equ:auc}
	AOC = Sensitivity(TPR)-(1-Specificity)(FPR) Curve
\end{equation}

\bibliographystyle{splncs04}